# SCF framework, HF stability and RPA correlation for Jordan–Wigner-transformed spin Hamiltonians on arbitrary coupling topologies

Shadan Ghassemi Tabrizi,[1,2]* Thomas M. Henderson,[3,4] Thomas D. Kühne,[1,2]

and Gustavo E. Scuseria[3,4]

[1]*Computational System Sciences, Technische Universität Dresden, 01187 Dresden, Germany*

[2]*Center for Advanced Systems Understanding (CASUS), Am Untermarkt 20, 02826 Görlitz, Germany, *s.ghassemi-tabrizi@hzdr.de*

[3]*Department of Physics and Astronomy, Rice University, Houston, TX 77005-1892, USA*

[4]*Department of Chemistry, Rice University, Houston, TX 77005-1892, USA*

**Abstract.** Mapping spins to fermions via the Jordan–Wigner (JW) transformation can render mean-field (Hartree–Fock, HF) descriptions effective for strongly correlated spin systems. As established in recent work, the application of such approaches is not limited by the nonlocal structure of JW strings or by site ordering, because string operators can be absorbed into Thouless rotations of a Slater determinant, and the variational optimization of a unitary Lie-Algebraic similarity transformation removes any ordering dependence. Leveraging these ideas, we develop a self-consistent field (SCF) scheme that expresses the mean-field energy as a functional of the single-particle density matrix, providing an alternative to gradient-based optimization of Thouless parameters. We derive the analytic orbital Hessian to diagnose HF stability and compute ground-state correlation energy through the random-phase approximation (RPA). Benchmark results for the *XXZ* and $J_1 - J_2$ model on one- and two-dimensional lattices demonstrate that RPA significantly improves mean-field accuracy.

## 1. Introduction

The physical properties of exchange-coupled spin clusters (e.g., magnetic molecules or lattice fragments of quantum magnets) are commonly modeled with spin Hamiltonians[1,2] such as the Heisenberg model, $H = \sum_{m<n} J_{mn} \mathbf{s}_m \cdot \mathbf{s}_n$. Although the use of spin and point-group symmetries to factor the Hamiltonian into invariant subspaces[3–5] extends the reach of exact diagonalization (ED), the exponential growth of the Hilbert space with the number of sites still confines exact calculations to relatively small lattices. A wide range of approximation schemes – e.g., density-matrix renormalization group[6,7] (DMRG) and other tensor-network methods,[8] quantum Monte Carlo,[9] coupled-cluster theory,[10] or spin-wave expansions[11] – address complementary regimes of quantum magnets with different computational costs, accuracies, and accessible observables. A comprehensive review is beyond the scope of this work.

Against this landscape, mean-field ideas[12–14] remain attractive for their conceptual transparency, low computational cost, and broad applicability. In earlier work[14] we implemented efficient mean-field treatments of JW-transformed spin Hamiltonians and captured correlation via Lie-Algebraic similarity transformations[13,15] (LAST). Here we recast the Hartree–Fock (HF) problem in an SCF formulation familiar from quantum chemistry. We express the mean-field energy as a functional of the single-particle density matrix $\rho$, interpret JW strings as Thouless rotations,[12] and remove site-ordering dependence via unitary LAST (combined with HF, this constitutes the orbital-optimized uLAST method,[13] oo-uLAST, see Theory section). The SCF formulation allows to leverage standard convergence machinery (damping, level shifting, DIIS, etc., see, e.g., Ref. 16 and references cited therein), provides stability diagnostics via the analytic orbital Hessian[17–19] derived here, and interfaces naturally with response theory, specifically, the random-phase approximation (RPA) for correlation energies[20–23] on top of the oo-uLAST reference. This yields a scalable mean-field framework for ground states of spin Hamiltonians with arbitrary coupling topologies, complementing the nonunitary LAST correlation approach from recent work.[13,14]

In the following Theory section, we first briefly recapitulate the JW transformation,[24] with emphasis on its extended variant[25,26] (EJW). The uLAST ansatz amounts to a variational optimization of the EJW parameters.[13] We then formulate the energy of a Slater determinant in the EJW fermionic representation as a functional of $\rho$ and by deriving the associated Fock matrix to be employed in an SCF scheme. Next, we construct the orbital Hessian and the closely related RPA matrix. Their spectra provide stability diagnostics of stationary HF points (distinguishing minima from saddle points) and access to ground-state correlation energies with respect to oo-uLAST, where we employ the ring-CCD convention[21,27] (prefactor $\frac{1}{4}$ in Eq. (38) below) for exchange-including kernels.

The Results section examines a set of one- and two-dimensional lattices, including various parameter regimes of the *XXZ* and antiferromagnetic $J_1 - J_2$ models. The Appendix derives the Fock kernel required to construct the orbital-Hessian and RPA matrices and provides an alternative formulation of the uLAST gradient.

## 2. Theory

**Extended JW and uLAST.** The (extended) JW mapping of spins to fermions,

$$s_p^+ = c_p^\dagger \phi_p^\dagger ,\tag{1}$$

$$s_p^- = c_p \phi_p ,\tag{2}$$

$$s_p^z = n_p - \tfrac{1}{2} ,\tag{3}$$

expresses the spin ladder operators at site $p$, $s_p^+ = s_p^x + i s_p^y$ and $s_p^- = s_p^x - i s_p^y$ in Eqs. (1) and (2), in terms of fermionic creation and annihilation operators, $c_p^\dagger$ and $c_p$. Up to a constant shift, the *z*-component of the local spin, $s_p^z$, corresponds to the occupation number of the respective orbital, $n_p = c_p^\dagger c_p$. The correct commutation (anticommutation) relations for spins (fermions) are afforded by string operators, Eq. (4), parametrized by real angles $\theta_{pq}$, under the constraints $\theta_{pq} = \theta_{qp} + \pi$ for $p < q$, and $\theta_{pp} = 0$.

$$\phi_p^\dagger = e^{i \sum_q \theta_{pq} n_q}\tag{4}$$

The conventional JW transformation is a special case in which $\theta_{pq} = 0$ for $q < p$, and we denote the corresponding fermionic Hamiltonian by $H_{\text{JW}}$. The more general $H_{\text{EJW}}$ is obtained by a unitary Lie-algebraic similarity transformation (uLAST) of $H_{\text{JW}}$, Eq. (5),

$$H_{\text{EJW}} = e^{-\gamma} H_{\text{JW}} e^{\gamma} ,\tag{5}$$

by the two-body correlator of Eq. (6):

$$\gamma = \tfrac{1}{2} \sum_{p,q} \gamma_{pq} n_p n_q ,\tag{6}$$

where $\gamma_{pq} = i\Theta_{pq}$, with a real symmetric matrix $\boldsymbol{\Theta}$; the strings occurring in $H_{\text{EJW}}$ (cf. Eq. (4)) are then defined by angles $\theta_{pq} = \Theta_{pq}$ for $p < q$ and $\theta_{pq} = \Theta_{pq} + \pi$ for $p > q$. Treating EJW-angles $\theta_{pq}$ as optimization parameters makes the uLAST ansatz independent of the site numbering.

**SCF formulation.** Within the oo-uLAST framework, we variationally optimize both the free EJW parameters (corresponding to a uLAST approach) and a Slater determinant for the EJW-transformed Hamiltonian (this HF part represents the orbital optimization). In our recent work, we showed how to optimize uLAST parameters by gradient descent simultaneously with orbital optimization.[14] In the Appendix of the present paper, we derive an alternative uLAST gradient evaluated at an HF stationary point, which is expressed in terms of $\boldsymbol{\rho}$ and does not reference the HF wave function. Together with the SCF formulation introduced below, this density-matrix form could support schemes akin to finite-temperature (thermal) HF,[28] though we do not pursue that direction here.

In what follows we cast the HF problem as an SCF procedure. With properly defined values for one- and two-particle integrals, $t_{kl}$ and $[kn|lm]$, the $z$-coupling contains no strings and can be written in the standard second-quantized form of Eq. (7), with a constant contribution $E_{\text{const}} = \frac{1}{4}\sum_{m<n} J_{mn}$.

$$\sum_{m<n} J_{mn} s_m^z s_n^z = E_{\text{const}} + \sum_{kl} t_{kl} c_k^\dagger c_l + \frac{1}{2}\sum_{klmn}[kn|lm] c_k^\dagger c_l^\dagger c_m c_n \qquad (7)$$

Thus a Fock matrix $\mathbf{F}^z$ for $z$-coupling can be constructed from the single-particle density matrix, $\rho_{kl} \equiv \langle\Phi|c_l^\dagger c_k|\Phi\rangle$, in the usual way, see Eq. (8),

$$\mathbf{F}^z = \mathbf{t} + \boldsymbol{\Gamma}, \qquad (8)$$

with $\boldsymbol{\Gamma}$ defined in Eq. (9):

$$\Gamma_{kl} = \sum_{mn}[kl|mn]\rho_{nm} . \qquad (9)$$

The respective contribution to the mean-field energy is given in Eq. (10):

$$E^z[\boldsymbol{\rho}] = \text{Tr}(\mathbf{t}\boldsymbol{\rho}) + \tfrac{1}{2}\text{Tr}(\boldsymbol{\Gamma}\boldsymbol{\rho}) + E_{\text{const}} . \qquad (10)$$

In contrast, $xy$-coupling involves string operators (strings vanish in open chains with nearest-neighbor interactions; in rings, a remaining string for the coupling between the first and last site reduces to a sign factor in a definite fermion-number sector). For a pair $\langle m,n\rangle$, we can pull the string operators to the right, Eq. (11),

$$s_m^x s_n^x + s_m^y s_n^y = \tfrac{1}{2} s_m^+ s_n^- + \text{h.c.} = \tfrac{1}{2}\left[c_m^\dagger c_n \phi_{m(n)}^\dagger \phi_{n(m)} + \text{h.c.}\right], \tag{11}$$

by defining reduced strings, $\phi_{m(n)}^\dagger = e^{i\sum_{q\neq n}\theta_{mq}n_q}$. The insight that strings act on Slater determinants as Thouless rotations[12] has recently enabled an efficient implementation of mean-field and subsequent correlation methods.[14] The unitary orbital rotation $\mathbf{R} \in \mathbb{C}^{N_{orb}\times N_{orb}}$ effected by a general string $e^{i\alpha_q n_q}$ is given in Eq. 1, where $\mathbf{C} \in \mathbb{C}^{N_{orb}\times N_f}$ and $\mathbf{C}_R$ collect the orthonormal occupied orbitals defining the Slater determinants $|\Phi\rangle$ and $|\Phi_R\rangle \equiv e^{i\sum_q \alpha_q n_q}$, respectively; $N_{orb}$ is the size of the single-particle basis, and $N_f$ is the fermion number.

$$\mathbf{C}_R = \mathbf{RC} = \begin{pmatrix} e^{i\alpha_1} & 0 & 0 \\ 0 & \ddots & 0 \\ 0 & 0 & e^{i\alpha_{N_{orb}}} \end{pmatrix} \mathbf{C} \tag{12}$$

With the aim of deriving the *xy*-coupling contribution to the Fock operator, $\mathbf{F}^{xy}$, we now formulate the respective mean-field energy as a functional of the single-particle density matrix, $\boldsymbol{\rho} = \mathbf{CC}^\dagger$. Eq. (13) for the overlap between the original and the rotated determinant invokes the overlap matrix $\mathbf{S} \equiv \mathbf{C}^\dagger \mathbf{C}_R$.[29]

$$w \equiv \langle \Phi | \Phi_R \rangle = \det(\mathbf{S}) \tag{13}$$

To express $w$ as a functional of $\boldsymbol{\rho}$, we introduce the auxiliary matrix $\mathbf{Z}$ in Eq. (14), where $\mathbf{1}$ is the unit matrix.

$$\mathbf{Z} = \mathbf{1} + \boldsymbol{\rho}(\mathbf{R} - \mathbf{1}) \tag{14}$$

Using the matrix-determinant identity of Eq. (15),

$$\det(\mathbf{1} + \mathbf{LM}) = \det(\mathbf{1} + \mathbf{ML}), \tag{15}$$

and setting $\mathbf{L} = \mathbf{C}$ and $\mathbf{M} = \mathbf{C}^\dagger(\mathbf{R} - \mathbf{1})$, we obtain Eq. (16):

$$\det(\mathbf{Z}) = \det[\mathbf{1} + \mathbf{CC}^\dagger(\mathbf{R} - \mathbf{1})] = \det[\mathbf{1} + \mathbf{C}^\dagger(\mathbf{R} - \mathbf{1})\mathbf{C}] = \det(\mathbf{S}) = w. \tag{16}$$

The single-particle transition-density matrix $\boldsymbol{\rho}_R$ is defined in Eq. (17):

$$(\boldsymbol{\rho}_R)_{kl} = \frac{\langle \Phi | c_l^\dagger c_k | \Phi_R \rangle}{\langle \Phi | \Phi_R \rangle}. \tag{17}$$

Eq. (18) is a standard result from a generalized Wick's theorem:

$$\boldsymbol{\rho}_R = \mathbf{R}\mathbf{C}\mathbf{S}^{-1}\mathbf{C}^\dagger . \tag{18}$$

In order to recast $\boldsymbol{\rho}_R$ as a density-matrix functional, we write Eq. (19),

$$\begin{aligned}\mathbf{Z}\mathbf{C}\mathbf{S}^{-1}\mathbf{C}^\dagger &= (\mathbf{1} + \mathbf{C}\mathbf{C}^\dagger\mathbf{R} - \mathbf{C}\mathbf{C}^\dagger)\mathbf{C}\mathbf{S}^{-1}\mathbf{C}^\dagger = \\ &= \mathbf{C}\mathbf{C}^\dagger\mathbf{R}\mathbf{C}\mathbf{S}^{-1}\mathbf{C}^\dagger + (\mathbf{1} - \mathbf{C}\mathbf{C}^\dagger)\mathbf{C}\mathbf{S}^{-1}\mathbf{C}^\dagger = \boldsymbol{\rho} + 0\end{aligned}, \tag{19}$$

hence:

$$\mathbf{Z}^{-1}\boldsymbol{\rho} = \mathbf{C}\mathbf{S}^{-1}\mathbf{C}^\dagger . \tag{20}$$

Combining Eqs. (20) and (18) then yields Eq. (21):

$$\boldsymbol{\rho}_R = \mathbf{R}\mathbf{Z}^{-1}\boldsymbol{\rho} . \tag{21}$$

Eq. (22) gives the energy contribution from $xy$-coupling for a pair $\langle m,n \rangle$,

$$E_{mn}^{xy} = \langle \Phi | s_m^x s_n^x + s_m^y s_n^y | \Phi \rangle = \tfrac{1}{2}\langle \Phi | c_m^\dagger c_n \phi_{m(n)}^\dagger \phi_{n(m)} + \text{h.c.} | \Phi \rangle = w_{mn}\text{Tr}(\mathbf{h}_{mn}\boldsymbol{\rho}_R^{mn}) + \text{c.c.} , \tag{22}$$

where $(\mathbf{h}_{mn})_{ij} = \tfrac{1}{2}J_{mn}\delta_{im}\delta_{jn}$; $w_{mn}$ and $\boldsymbol{\rho}_R^{mn}$ are calculated with respect to the rotated Slater determinant $|\Phi_R\rangle = \phi_{m(n)}^\dagger \phi_{n(m)}|\Phi\rangle$.

The Fock matrix represents the functional derivative $\mathbf{F} = \delta E / \delta \boldsymbol{\rho}$; in terms of differentials, $dE = \text{Tr}(\mathbf{F}d\boldsymbol{\rho})$. $\mathbf{F}$ is a sum of contributions from $z$-coupling, Eq. (8), and $xy$-coupling, $\mathbf{F} = \mathbf{F}^z + \mathbf{F}^{xy}$. To obtain $\mathbf{F}^{xy}$, we consider the first term on the right-hand side of Eq. (22), i.e., the incremental energy contribution $\mathcal{E}_{mn} = w_{mn}\text{Tr}(\mathbf{h}_{mn}\boldsymbol{\rho}_R^{mn})$. We shall extract the respective Fock increment $\mathbf{f}_{mn}$ from $d\mathcal{E}_{mn} = \text{Tr}(\mathbf{f}_{mn}d\boldsymbol{\rho})$. In the following derivation – leading to Eq. (28) – to avoid clutter, we write $\mathcal{E}$, $w$, $\mathbf{h}$, $\boldsymbol{\rho}_R$ and $\mathbf{f}$ instead of $\mathcal{E}_{mn}$, $w_{mn}$, $\mathbf{h}_{mn}$, $\boldsymbol{\rho}_R^{mn}$ and $\mathbf{f}_{mn}$, respectively. The differentials for $\mathbf{Z}$ and $\mathbf{Z}^{-1}$ are given in Eqs. (23) and (24), respectively,

$$d\mathbf{Z} = (d\boldsymbol{\rho})(\mathbf{R} - \mathbf{1}) , \tag{23}$$

$$d(\mathbf{Z}^{-1}) = -\mathbf{Z}^{-1}(d\mathbf{Z})\mathbf{Z}^{-1} = -\mathbf{Z}^{-1}(d\boldsymbol{\rho})(\mathbf{R} - \mathbf{1})\mathbf{Z}^{-1} , \tag{24}$$

thus:

$$d\boldsymbol{\rho}_R = \mathbf{R}d(\mathbf{Z}^{-1})\boldsymbol{\rho} + \mathbf{R}\mathbf{Z}^{-1}d\boldsymbol{\rho} = -\mathbf{R}\mathbf{Z}^{-1}(d\boldsymbol{\rho})(\mathbf{R} - \mathbf{1})\mathbf{Z}^{-1}\boldsymbol{\rho} + \mathbf{R}\mathbf{Z}^{-1}d\boldsymbol{\rho} . \tag{25}$$

For the differential of the overlap, Eq. (26), Jacobi's determinant formula and Eq. (23) are employed,

$$dw = d\det(\mathbf{Z}) = \det(\mathbf{Z})\text{Tr}(\mathbf{Z}^{-1}d\mathbf{Z}) = w\text{Tr}[\mathbf{Z}^{-1}(d\boldsymbol{\rho})(\mathbf{R}-\mathbf{1})] \ , \qquad (26)$$

which yields the energy differential in Eq. (27).

$$\begin{aligned}d\mathcal{E} &= w\text{Tr}(\mathbf{h}d\boldsymbol{\rho}_R) + dw\text{Tr}(\mathbf{h}\boldsymbol{\rho}_R) = \\ &w\text{Tr}(\mathbf{hRZ}^{-1}d\boldsymbol{\rho}) - w\text{Tr}[\mathbf{hRZ}^{-1}(d\boldsymbol{\rho})(\mathbf{R}-\mathbf{1})\mathbf{Z}^{-1}\boldsymbol{\rho}] + w\text{Tr}[\mathbf{Z}^{-1}(d\boldsymbol{\rho})(\mathbf{R}-\mathbf{1})]\text{Tr}(\mathbf{h}\boldsymbol{\rho}_R)\end{aligned} \qquad (27)$$

Using the cyclicity of the trace to collect every occurrence of $d\boldsymbol{\rho}$ on the right, we finally obtain the Fock increment, Eq. (28),

$$\mathbf{f} = w[\mathbf{hRZ}^{-1} - (\mathbf{R}-\mathbf{1})\mathbf{Z}^{-1}\boldsymbol{\rho}\mathbf{hRZ}^{-1} + \text{Tr}(\mathbf{h}\boldsymbol{\rho}_R)(\mathbf{R}-\mathbf{1})\mathbf{Z}^{-1}] \ . \qquad (28)$$

$\mathbf{F}^{xy}$ represents the sum of increments for all ordered pairs, Eq. (29),

$$\mathbf{F}^{xy} = \sum_{m\neq n}\mathbf{f}_{mn} \ . \qquad (29)$$

The usual SCF process consists in the following steps: i) build the Fock matrix $\mathbf{F} = \mathbf{F}^z + \mathbf{F}^{xy}$ for some initial guess $\boldsymbol{\rho}$, ii) diagonalize $\mathbf{F}$, i.e., $\mathbf{F}\mathbf{v}_i = \varepsilon_i\mathbf{v}_i$, yielding orbital energies $\varepsilon_i$ and canonical orbitals $\mathbf{v}_i$ (molecular orbitals, MOs, in quantum-chemical terminology), iii) occupy the orbitals following the aufbau principle to form a new density matrix, $\boldsymbol{\rho} = \sum_{p=1}^{N_f}\mathbf{v}_p\mathbf{v}_p^\dagger$, iv) repeat until self-consistency is reached, where $[\mathbf{F},\boldsymbol{\rho}]=0$. Standard techniques for SCF acceleration can be employed. Note that even if one proceeds with a Thouless-parameter optimization rather than an SCF process, access to canonical orbitals and orbital energies is still valuable because it supplies a diagonal preconditioner for the gradient. As an example, rescaling the virtual-occupied blocks by $(\Delta\varepsilon_{vo})^{-1/2}$, where $\Delta\varepsilon_{vo} = \varepsilon_v - \varepsilon_o$ is the respective orbital-energy difference, was suggested in Ref. 30 to accelerate convergence.

The dominant cost of mean-field calculations on general coupling graphs comes from the $xy$-coupling terms, whose strings act as orbital rotations on a Slater determinant, see Eq. (12). As explained, evaluating the energy/Fock contributions for an interacting pair $\langle m,n\rangle$ requires overlaps and transition densities between $|\Phi\rangle$ and a string-rotated determinant $|\Phi_R\rangle$. The formal leading cost of building $\mathbf{f}_{mn}$ scales as $\mathcal{O}(N_{\text{orb}}^3)$, with $N_{\text{orb}}$ being the number of spin-1/2 sites (or spin-1/2 auxiliaries representing $s > \frac{1}{2}$ sites, see below). The number of interacting pairs is typically $\mathcal{O}(N_{orb})$, implying an overall $\mathcal{O}(N_{orb}^4)$ scaling for a full Fock build; the subsequent diagonalization of **F** scales as $\mathcal{O}(N_{orb}^3)$. In the Thouless-parameter optimization route,[14] the mean-field state is parametrized by complex amplitudes $Z_{vo}$. Each optimizer step requires an evaluation of the energy and its analytic gradient, which involves the same string-rotated overlaps/transition densities as above, hence the same $\mathcal{O}(N_{orb}^4)$ leading cost; an additional transformation from local to global gradient representations is an $\mathcal{O}(N_{orb}^3)$ step in the standard implementation.

We optimize the mean-field state and the uLAST parameters in a coupled fashion using a nested (macro/micro) procedure. In each macro-iteration, the mean-field state is updated for fixed uLAST parameters by solving the SCF equations to (near-)stationarity, after which the uLAST parameters are updated using a gradient-based step evaluated at the (approximately) stationary mean-field solution. We accelerate SCF convergence using the OpenOrbitalOptimizer (OOO) package[16] of Lehtola and Burns, a lightweight SCF driver providing standard stabilization and convergence-acceleration schemes (e.g., damping/level shifting and DIIS-type extrapolation). OOO is connected to our MATLAB code through a small C++/MEX interface layer that implements a callback: given the current orbital coefficients and occupations proposed by OOO, the interface constructs the (generally complex) density matrix $\boldsymbol{\rho} = \mathbf{CC}^\dagger$ and calls a MATLAB routine to evaluate the HF energy, $E[\boldsymbol{\rho}]$, and the Fock matrix, $\mathbf{F}[\boldsymbol{\rho}]$. These quantities are returned to OOO, which carries out the diagonalization and updates the orbitals using its built-in acceleration strategy. This cycle is repeated until self-consistency is reached. For robustness, we first perform a short preconditioning stage with our original MATLAB SCF loop using simple density damping and then start OOO from the corresponding Fock matrix.

**Gauge freedom in oo-uLAST.** We note a redundancy in the parametrization of the real symmetric matrix $\boldsymbol{\Theta}$ that defines the uLAST correlator $\gamma$ (cf. Eq. (6)). Consider the separable shift of Eq. (30),

$$\Theta_{pq} \to \Theta'_{pq} = \Theta_{pq} + \chi_p + \chi_q, \quad \chi_p \in \mathbb{R}, \tag{30}$$

which changes the generator $\Delta\gamma = \gamma(\boldsymbol{\Theta}') - \gamma(\boldsymbol{\Theta})$,

$$\Delta\gamma = \frac{i}{2}\sum_{p<q}(\chi_p + \chi_q)n_p n_q = \frac{i}{2}\sum_p \chi_p n_p \sum_{q \neq p} n_q = \frac{i}{2}\sum_p \chi_p n_p (\mathcal{N}-1), \tag{31}$$

where $\mathcal{N} \equiv \sum_q n_q$. Within a fixed particle-number sector, $\mathcal{N} = N_f$, Eq. (31) reduces to a purely one-body operator, Eq. (32),

$$\Delta \gamma = i \sum_p \kappa_p n_p, \quad \kappa_p = \tfrac{1}{2}(N_f - 1)\chi_p . \tag{32}$$

Since $\gamma(\Theta)$ and $\Delta\gamma$ commute, $[\gamma(\Theta), \Delta\gamma] = 0$, the shifted Hamiltonian is related to the original one by a unitary similarity transformation, Eq. (33),

$$H_{\text{EJW}}(\Theta') = e^{-\gamma(\Theta')} H_{\text{JW}} e^{\gamma(\Theta')} = U_\chi^\dagger H_{\text{EJW}}(\Theta) U_\chi , \tag{33}$$

where $U_\chi \equiv e^{\Delta\gamma}$ represents a local U(1) transformation of the fermionic modes, Eqs. (34) and (35):

$$U_\chi^\dagger c_p U_\chi = e^{i\kappa_p} c_p , \tag{34}$$

$$U_\chi^\dagger c_p^\dagger U_\chi = e^{-i\kappa_p} c_p^\dagger . \tag{35}$$

Thus, within a fixed particle-number sector, the shift in Eq. (30) leaves the spectrum and all number-conserving observables invariant. This is a gauge redundancy of the parametrization that manifests as flat directions in the $\Theta$ parameter space. A simple gauge fixing is obtained by choosing a reference index $r$ and imposing $\Theta_{rq} = \Theta_{qr} = 0$ for all $q$, which removes $N_{\text{orb}} - 1$ redundant degrees of freedom and thereby reduces the number of independent uLAST parameters from $\tfrac{1}{2} N_{\text{orb}}(N_{\text{orb}} - 1)$ to $\tfrac{1}{2}(N_{\text{orb}} - 1)(N_{\text{orb}} - 2)$. However, we found that the described gauge fixing frequently worsened the convergence properties of oo-uLAST calculations, and we therefore retain the redundant parametrization in the production calculations.

**HF stability and RPA.** To assess the character of a stationary HF solution, we analyze the orbital Hessian, which defines the second derivative of the HF energy with respect to particle-hole (Thouless) rotations, see Eq. (65) below. In the particle-hole basis it has block form:

$$\mathbf{H} = \begin{pmatrix} \mathbf{A} & \mathbf{B} \\ \mathbf{B}^* & \mathbf{A}^* \end{pmatrix} . \tag{36}$$

A positive-definite $\mathbf{H}$ certifies HF stability[18,31] (a local minimum), broken continuous one-body symmetries cause zero modes, and any negative eigenvalue signals an instability (a corresponding eigenvector provides a direction that can be used to distort the Slater determinant toward lower energy[32]). RPA, on the other hand, leads to an eigenproblem for the matrix **R** of Eq. (37),

$$\mathbf{R} = \begin{pmatrix} \mathbf{A} & \mathbf{B} \\ -\mathbf{B}^* & -\mathbf{A}^* \end{pmatrix}, \tag{37}$$

which is constructed from the same blocks but with a different metric.[33,34] For a positive-definite orbital Hessian, the RPA spectrum is real (conversely, a real RPA spectrum is not sufficient to guarantee positive-definiteness of the orbital Hessian[35,36]).

For a stable HF solution, the eigenvalues of **R** occur in real pairs $\pm\omega$, and these RPA eigenmodes can be used to estimate ground-state correlation energies. We evaluate the RPA correlation energy $E_c^{\text{RPA}}$ in the ring-CCD convention, Eq. (38), for a oo-uLAST reference via the summed difference between the spectra of RPA and TDA (Tamm-Dancoff approximation, setting $\mathbf{B} = 0$ in the RPA matrix, eigenvalues $\pm\nu$), considering only the positive RPA and TDA eigenvalues:

$$E_c^{\text{RPA}} = \frac{1}{4}\sum_i (\omega_i - \nu_i) = \frac{1}{4}\left[\sum_i \omega_i - \text{Tr}(\mathbf{A})\right]. \tag{38}$$

This choice is consistent with interpreting our exchange-including JW kernel as a fermionic ring-CCD resummation,[21] whereas the plasmon formula[20] would assign a prefactor of $\frac{1}{2}$ to the same spectrum, thus yielding correlation energies larger by a factor of two.

Due to the nonlocal JW string operators, the fermionic Hamiltonian is not restricted to the usual one- and two-body terms. As a result, existing derivations of the orbital Hessian or the RPA matrix[33,34] may not transfer directly to this setting. In the following, we therefore provide a self-contained, transparent derivation of **A** and **B** explicitly from the Fock kernel **K** (the derivative of **F** with respect to $\boldsymbol{\rho}$; see Appendix for a complete derivation of **K**).

A stationary point (HF solution) $\boldsymbol{\rho}_0$ satisfies $[\mathbf{F}_0, \boldsymbol{\rho}_0]$, with $\mathbf{F}_0 \equiv \mathbf{F}[\boldsymbol{\rho}_0]$. We work in the canonical MO basis, where $\mathbf{F}_0$ and $\boldsymbol{\rho}_0$ are diagonal, i.e., $\mathbf{F}_0 = \text{diag}(\{\varepsilon_p\}) \equiv \boldsymbol{\varepsilon}$ (orbital energies are sorted in ascending order) and $\boldsymbol{\rho}_0 = \text{diag}(\mathbf{1}_{N_o}, \mathbf{0}_{N_v})$; $N_o$ and $N_v$ are the numbers of occupied and virtual orbitals, respectively. According to the Thouless theorem,[17] every non-orthogonal Slater determinant – and thus every nearby Slater determinant – results from an occupied-virtual unitary transformation (Thouless rotation) of $\boldsymbol{\rho}_0$, Eq. (39),

$$\boldsymbol{\rho} = e^{i\boldsymbol{\kappa}} \boldsymbol{\rho}_0 e^{-i\boldsymbol{\kappa}}, \tag{39}$$

where $t$ is a real scaling parameter, and the anti-Hermitian $\boldsymbol{\kappa} = -\boldsymbol{\kappa}^\dagger$, Eq. (40),

$$\boldsymbol{\kappa} = \mathbf{T} - \mathbf{T}^\dagger , \qquad (40)$$

is defined in terms of a matrix $\mathbf{T}$, Eq. (41),

$$\mathbf{T} = \begin{pmatrix} \mathbf{0} & \mathbf{0} \\ \mathbf{Z} & \mathbf{0} \end{pmatrix} , \qquad (41)$$

comprising occupied-virtual transition amplitudes (Thouless parameters), $\mathbf{Z} \in \mathbb{C}^{N_v \times N_o}$. In operator form, Eq. (42),

$$T = \sum_{vo} Z_{vo} a_v^\dagger a_o , \qquad (42)$$

$a_p^\dagger$ ($a_p$) creates (annihilates) a fermion in the $p$-th MO, see Eq. (43):

$$a_p^\dagger = \sum_q C_{qp} c_q^\dagger . \qquad (43)$$

In the Taylor expansion of Eq. (44),

$$\boldsymbol{\rho}(t) = \boldsymbol{\rho}_0 + \boldsymbol{\rho}_0' t + \frac{1}{2} \boldsymbol{\rho}_0'' t^2 + \mathcal{O}(t^3) , \qquad (44)$$

the first- and second-order derivatives, Eqs. (45) and (46), result from the Baker-Campbell-Hausdorff expansion.

$$\left.\frac{d\boldsymbol{\rho}}{dt}\right|_{t=0} \equiv \boldsymbol{\rho}_0' = [\boldsymbol{\kappa}, \boldsymbol{\rho}_0] = \begin{pmatrix} \mathbf{0} & \mathbf{Z}^\dagger \\ \mathbf{Z} & \mathbf{0} \end{pmatrix} \qquad (45)$$

$$\boldsymbol{\rho}_0'' = [\boldsymbol{\kappa},[\boldsymbol{\kappa},\boldsymbol{\rho}_0]] = 2\begin{pmatrix} -\mathbf{Z}^\dagger \mathbf{Z} & \mathbf{0} \\ \mathbf{0} & \mathbf{Z}\mathbf{Z}^\dagger \end{pmatrix} \qquad (46)$$

In the second-order expansion of the HF energy, with $E_0 \equiv E[\boldsymbol{\rho}_0]$, Eq. (47),

$$E[\boldsymbol{\rho}] = E_0 + E_0' t + \frac{1}{2} E_0'' t^2 + \mathcal{O}(t^3) , \qquad (47)$$

the first derivative, Eq. (48),

$$E' = \text{Tr}(\mathbf{F}\boldsymbol{\rho}') , \qquad (48)$$

vanishes at the stationary point, Eq. (49):

$$E_0' = \text{Tr}(\mathbf{F}_0 \boldsymbol{\rho}_0') = \text{Tr}(\mathbf{F}_0 [\boldsymbol{\kappa},\boldsymbol{\rho}_0]) = \text{Tr}([\mathbf{F}_0,\boldsymbol{\rho}_0]\boldsymbol{\kappa}) = 0 . \qquad (49)$$

The second derivative is given by Eq. (50):

$$E'' = \mathrm{Tr}(\mathbf{F}'\boldsymbol{\rho}') + \mathrm{Tr}(\mathbf{F}\boldsymbol{\rho}'') \ . \tag{50}$$

Through the chain rule, the response of the Fock matrix to a density variation, Eq. (51),

$$F'_{pq} = \sum_{rs} K_{pqrs} \rho'_{rs} \ , \tag{51}$$

is encoded in the kernel **K**, Eq. (52),

$$K_{pqrs} \equiv \frac{\partial F_{pq}}{\partial \rho_{rs}} \ , \tag{52}$$

The first term in Eq. (50) can thereby be formulated as Eq. (53):

$$\mathrm{Tr}(\mathbf{F}'\boldsymbol{\rho}') = \sum_{pqrs} K_{pqrs} \rho'_{rs} \rho'_{qp} \ . \tag{53}$$

By the definition of the Fock matrix, $F_{pq} = \partial E / \partial \rho_{qp}$, the kernel collects the second energy derivatives, $K_{pqrs} = \partial^2 E / \partial \rho_{qp} \partial \rho_{rs}$ (note the order of indices in our convention). A derivation of **K** is provided in the Appendix.

The second term of Eq. (50) is evaluated at the HF point in Eq. (54), using the fact that $\mathbf{F}_0^{vv} = \boldsymbol{\varepsilon}_v$ and $\mathbf{F}_0^{oo} = \boldsymbol{\varepsilon}_o$ are diagonal.

$$\mathrm{Tr}[\mathbf{F}_0 \boldsymbol{\rho}''_0] = 2\left[\mathrm{Tr}(\mathbf{F}_0^{vv} \mathbf{Z}\mathbf{Z}^\dagger) - \mathrm{Tr}(\mathbf{F}_0^{oo} \mathbf{Z}^\dagger \mathbf{Z})\right] = 2\sum_{vo}(\varepsilon_v - \varepsilon_o)|Z_{vo}|^2 \tag{54}$$

Overall, the second energy derivative can be formulated as Eq. (55):

$$E''_0 = \sum_{pqrs} K_{pqrs} (\boldsymbol{\rho}'_0)_{rs} (\boldsymbol{\rho}'_0)_{qp} + 2\sum_{vo}(\varepsilon_v - \varepsilon_o)|Z_{vo}|^2 \ . \tag{55}$$

By vectorizing **Z** as $\mathbf{z} = \mathrm{vec}(\mathbf{Z}) \in \mathbb{C}^{N_v N_o}$ in column-major convention ($z_{v+(o-1)N_v} = Z_{vo}$; `Matlab` code: `z = Z(:)`), the orbital-gap term can be reformulated as in Eq. (56),

$$\sum_{v,o}(\varepsilon_v - \varepsilon_o)|Z_{vo}|^2 = \mathbf{z}^\dagger \cdot \mathbf{A}^\Delta \cdot \mathbf{z} \ , \tag{56}$$

with Eq. (57),

$$\mathbf{A}^\Delta = \mathbf{1}_{N_o} \otimes \boldsymbol{\varepsilon}_v - \boldsymbol{\varepsilon}_o \otimes \mathbf{1}_{N_v} \ , \tag{57}$$

or, in component notation:

$$A^{\Delta}_{ai,bj} = (\varepsilon_a - \varepsilon_i)\delta_{ab}\delta_{ij} . \tag{58}$$

In Eq. (58) and hereafter in this section, indices $a$, $b$ and $i$, $j$ are used for virtual and occupied MOs, respectively, and $p$, $q$, $r$, $s$ are used for arbitrary MOs.

With $\boldsymbol{\rho}'_0$ having only $vo$ and and $ov$ blocks, cf. Eq. (45), $(\boldsymbol{\rho}'_0)_{ai} = Z_{ai}$, $(\boldsymbol{\rho}'_0)_{ia} = Z^*_{ai}$, we reformulate the kernel term in Eq. (59):

$$\sum_{pqrs} K_{pqrs} (\boldsymbol{\rho}'_0)_{rs} (\boldsymbol{\rho}'_0)_{qp} =$$
$$\underbrace{\sum_{abij} K_{jbai} Z_{ai} Z_{bj}}_{(A)} + \underbrace{\sum_{abij} K_{bjai} Z_{ai} Z^*_{bj}}_{(B)} + \underbrace{\sum_{abij} K_{jbia} Z^*_{ai} Z_{bj}}_{(C)} + \underbrace{\sum_{abij} K_{bjia} Z^*_{ai} Z^*_{bj}}_{(D)} . \tag{59}$$

Using the general relations $K_{pqrs} = K^*_{rspq}$ and $K_{pqrs} = K^*_{qpsr}$, we have $K_{jbia} = K_{aibj}$ in term (C). Renaming indices $(i,a) \leftrightarrow (j,b)$ in term (B), $K_{bjai} Z_{ai} Z^*_{bj} \to K_{aibj} Z_{bj} Z^*_{ai}$, then shows that $(B) = (C)$. Similarly, we have $(A) = (D)^*$. Thus, by vectorizing $\mathbf{Z}$ as $\mathbf{z}$, we can write Eq. (60),

$$\sum_{pqrs} K_{pqrs} (\boldsymbol{\rho}'_0)_{rs} (\boldsymbol{\rho}'_0)_{qp} = 2\mathbf{z}^\dagger \cdot \mathbf{A}^K \cdot \mathbf{z} + (\mathbf{z}^\dagger \cdot \mathbf{B} \cdot \mathbf{z}^* + \text{c.c.}) , \tag{60}$$

with $\mathbf{A}^K$ and $\mathbf{B}$ defined in component form in Eqs. (61) and (62):

$$A^K_{ai,bj} = K_{aibj} , \tag{61}$$

$$B_{ai,bj} = K_{aijb} . \tag{62}$$

Putting this all together, we get Eq. (63):

$$E''_0 = 2\mathbf{z}^\dagger \cdot (\mathbf{A}^K + \mathbf{A}^\Delta) \cdot \mathbf{z} + (\mathbf{z}^\dagger \cdot \mathbf{B} \cdot \mathbf{z}^* + \text{c.c.}) . \tag{63}$$

Stacking the amplitudes and their conjugates into the vector $\mathbf{u}$ in Eq. (64),

$$\mathbf{u} = \begin{pmatrix} \mathbf{z} \\ \mathbf{z}^* \end{pmatrix} , \tag{64}$$

affords the standard orbital-Hessian block form, Eq. (65),

$$E''_0 = \mathbf{u}^\dagger \cdot \begin{pmatrix} \mathbf{A} & \mathbf{B} \\ \mathbf{B}^* & \mathbf{A}^* \end{pmatrix} \cdot \mathbf{u} = \mathbf{u}^\dagger \cdot \mathbf{H} \cdot \mathbf{u} , \tag{65}$$

where $\mathbf{A} = \mathbf{A}^K + \mathbf{A}^\Delta$. The analytic expression for the orbital Hessian was validated against numerical Hessians obtained from a Thouless-parameter optimization of the Slater determinant (see Ref. 14), using the `fminunc` function in the `MATLAB Optimization Toolbox`.

**Larger local spins.** To treat on-site spins $s > \frac{1}{2}$ within the JW framework, we represent each physical spin at site $p$ by $2s_p$ auxiliary spin-1/2 degrees of freedom $\{\kappa_{p,a}\}$, Eq. (66),

$$\mathbf{s}_p \to \sum_{a=1}^{2s_p} \kappa_{p,a} \ , \tag{66}$$

and apply the JW (or EJW) mapping to these auxiliaries. Although this enlarges the Hilbert space, because the $\{\kappa_{p,a}\}$ can couple to local spins smaller than $s_p$, the construction is variationally safe: as shown in our previous work,[14] for bilinear spin Hamiltonians, the exact ground state has maximal local spin at every site, and, when $S_z$ is conserved, the same holds within each $S_z$ sector (with definite magnetic quantum number $M$). Thus, any trial state in the enlarged space provides an upper bound to the exact ground-state energy, without requiring projection onto the maximal local-spin subspace. Adopting the auxiliary spin-1/2 representation allows us to use exactly the same methodology (oo-uLAST within the SCF formulation and the subsequent RPA treatment of correlation) for systems with $s > \frac{1}{2}$ sites, also with site-dependent $s$, although our examples focus on uniform local spin.

## 3. Results and Discussion

We assess the practical performance of evaluating the RPA correlation energy on top of oo-uLAST references based on the proposed SCF/HF framework for a small set of benchmark systems. Using exact diagonalization (ED) as a reference, we consider one- and two-dimensional coupling topologies in representative parameter regimes of the anisotropic *XXZ* model, Eq. (67),

$$H = \sum_{\langle m,n \rangle} \left( s_m^x s_n^x + s_m^y s_n^y + \Delta s_m^z s_n^z \right) \ , \tag{67}$$

and the isotropic $J_1 - J_2$ model, Eq. (68),

$$H = J_1 \sum_{\langle m,n \rangle} \mathbf{s}_m \cdot \mathbf{s}_n + J_2 \sum_{\langle\langle m,n \rangle\rangle} \mathbf{s}_m \cdot \mathbf{s}_n \ , \tag{68}$$

which includes antiferromagnetic nearest-neighbor (NN, $J_1 > 0$) and next-nearest-neighbor couplings (NNN, $J_2 > 0$). All RPA energies are computed with respect to the converged oo-uLAST reference in the $M = 0$ sector, corresponding to half-filling $N_f = N_{orb}/2$ in the fermionic picture. HF stability is assessed from the eigenvalue spectrum of the corresponding orbital Hessian, and we illustrate how the orbital-Hessian spectrum provides a diagnostic of the quality of the mean-field reference.

*XXZ* **model.** Figure 1a compares relative ground-state energy errors for an $N = 8$ spin-1/2 *XXZ* chain as a function of the anisotropy $\Delta$, using four levels of approximation: JW-HF, JW-HF+RPA, oo-uLAST, and oo-uLAST+RPA. Adding RPA correlation systematically improves energies in most of the parameter range, with the largest impact in regions where the underlying reference is least accurate. The orbital-Hessian spectrum for oo-uLAST plotted in Figure 1b explains the kink in the RPA energy curve at the isotropic point $\Delta = 1$ in terms of an eigenvalue approaching zero, indicating a change in the character of the stationary mean-field reference along the scan (i.e., competing solutions becoming nearly degenerate). Consistent with earlier observations for one-dimensional systems,[13,14] in the sequential numbering of sites along the chain, JW-HF and oo-uLAST are equivalent for $\Delta \geq 0$, whereas for $\Delta < 0$ oo-uLAST affords lower variational energies than JW-HF, and the corresponding oo-uLAST+RPA curve provides the best overall description in this regime. In the following, we restrict our discussion to oo-uLAST references, because the numbering dependence[13] and the performance of JW-HF for fixed site orderings relative to oo-uLAST have already been illustrated in earlier work.[13,14]

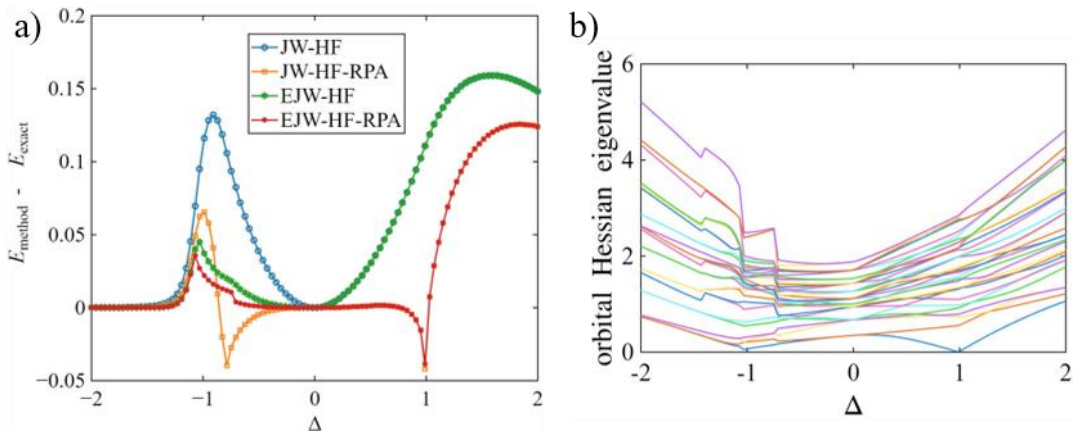

Figure 1: a) Relative ground-state energy errors for an open $N = 8$ spin-1/2 *XXZ* chain as a function of the anisotropy $\Delta$. Shown are results for HF (with sequential numbering of sites) and oo-uLAST. b) Spectrum of the orbital Hessian for oo-uLAST.

Figure 2 shows the performance of oo-uLAST and RPA in terms of the relative ground-state energy error, $(E_{\text{exact}} - E_{\text{method}})/E_{\text{exact}}$, for a family of systems with 24 spin-1/2 sites arranged in increasingly compact geometries: from a one-dimensional chain to quasi-2D ladders ($12 \times 2$, $8 \times 3$) and the more compact $6 \times 4$ layout, each considered with open and periodic boundary conditions (OBC/PBC). Across all geometries and boundary conditions, the RPA correlation reduces the energy error over essentially the entire $\Delta$ range. The improvement is particularly pronounced in regions where the oo-uLAST reference is least accurate – most visibly near the sharp feature around $\Delta \approx -1$ and in the broad maximum at positive $\Delta$ – and it persists as the connectivity becomes more two-dimensional. At the same time, the residual errors increase as the arrangement becomes more compact and when periodic boundary conditions are imposed. These results can be compared to our earlier LAST-based correlation treatment on the same set of systems.[14] While LAST typically achieves smaller absolute errors than RPA when it converges reliably, the present data show that RPA already captures a substantial fraction of the missing correlation at comparatively low additional cost and without the iterative amplitude optimization required by LAST. In this sense, oo-uLAST+RPA provides an efficient and robust correlation layer that complements the more demanding LAST treatment.

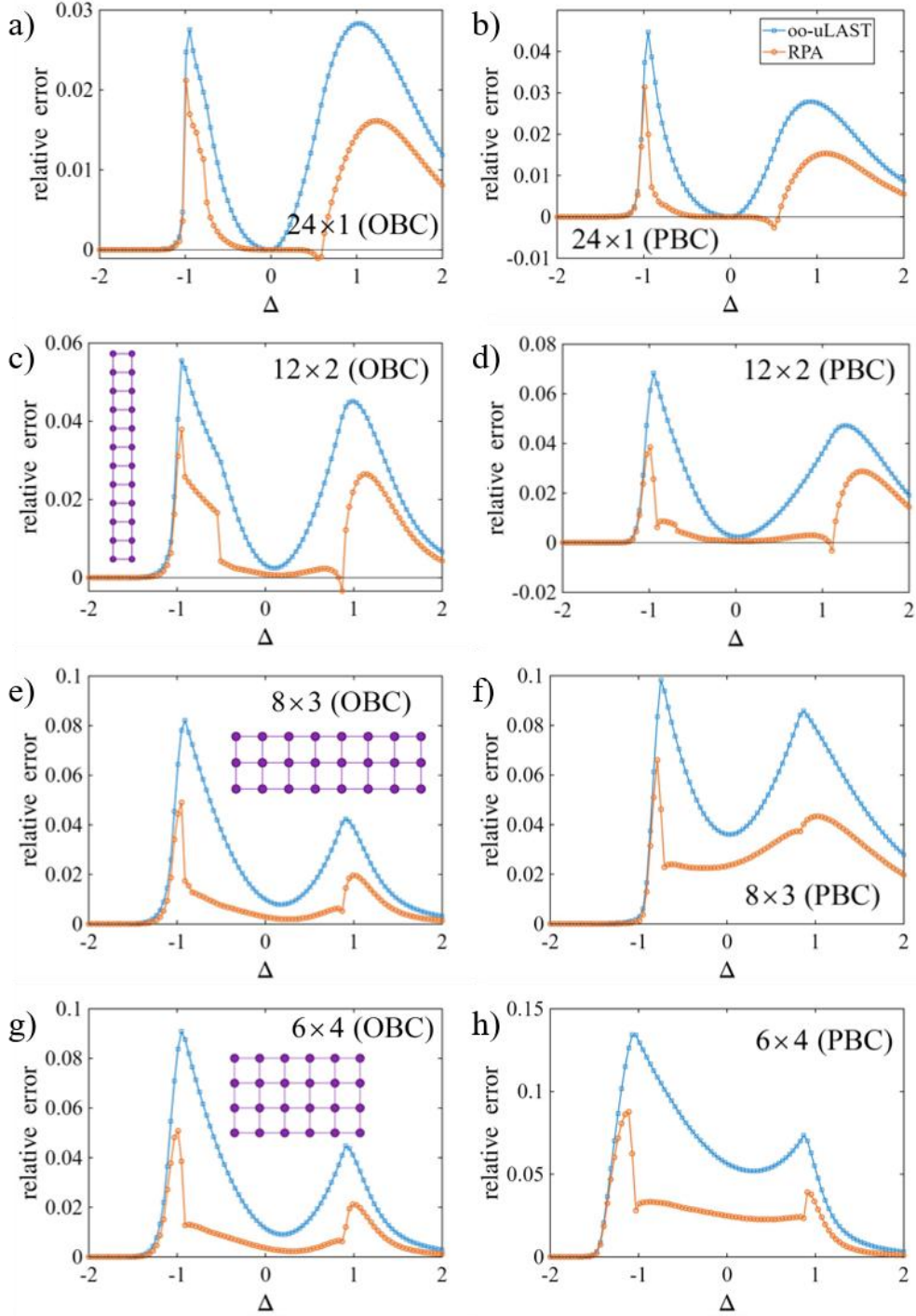

Figure 2: Relative ground-state energy errors for the spin- 1/2 *XXZ* model on 24-site lattices as a function of $\Delta$.

The oo-uLAST results shown in Figure 2 are identical to those reported in our earlier work,[14] and we have verified that the same solutions can be reproduced using the SCF algorithm introduced in the present work. For all systems, we verified that the final mean-field references are HF-stable according to the orbital-Hessian criterion. We attribute this in part to our protocol for selecting low-energy mean-field solutions along parameter scans:[14] for each $\Delta$ we converge multiple oo-uLAST optimizations from random initial guesses and then propagate solutions stepwise by seeding neighboring $\Delta$ values, retaining the lowest-energy solution at each point. Since JW-HF/oo-uLAST commonly admits several energetically close stationary solutions, this strategy increases the likelihood of tracking the lowest-energy (and typically stable) branch rather than converging to a nearby saddle point. Finally, note that different local optimization procedures for the HF problem (e.g., gradient-based minimization in Thouless parameters versus SCF iterations) can have different basins of attraction and therefore may converge to different stationary solutions even when started from the same initial guess.

*s > 1/2*. To illustrate how the approach performs beyond the spin-1/2 case, we consider an $N=4$ *XXZ* ring with increasing on-site spin *s*. For $s > \frac{1}{2}$, the EJW fermionization relies on auxiliary spin-1/2 degrees of freedom and therefore introduces nontrivial string structure even in one-dimensional systems. As a result, the mean-field description is not trivially exact in special limits (in contrast to the spin-1/2 case at $\Delta = 0$).

Figure 3 shows the relative ground-state energy error for oo-uLAST and RPA as a function of $\Delta$ in an $N=4$ ring of $s=1$ sites. The accuracy of oo-uLAST and its RPA correction correlate with the stability diagnostics provided by the orbital-Hessian spectrum. Over most of the scan, the RPA correction lowers the error substantially. A notable exception is the narrow interval around $\Delta \approx 1$, where both curves exhibit a sharp feature. Figure 3b helps rationalize this behavior: in the same region, the lowest orbital-Hessian eigenvalue approaches zero, indicating that the mean-field solution is soft with respect to orbital rotations. In addition, the Hessian eigenvalues show a kink, consistent with a near-degeneracy between mean-field solutions. Away from this narrow region, most clearly toward larger positive $\Delta$, both the mean-field and RPA-corrected energies approach the exact result as the ground state becomes increasingly classical in the easy-axis limit. Note that, even though the mean-field reference breaks the underlying $S_x$ and $S_y$ spin-rotation symmetry at the isotropic point, this does not produce orbital-Hessian zero modes (cf. Figure 3b), because rotations generated by $S_x$ and $S_y$ are not one-body symmetries in the JW/EJW fermionic representation (and they mix particle-number sectors).

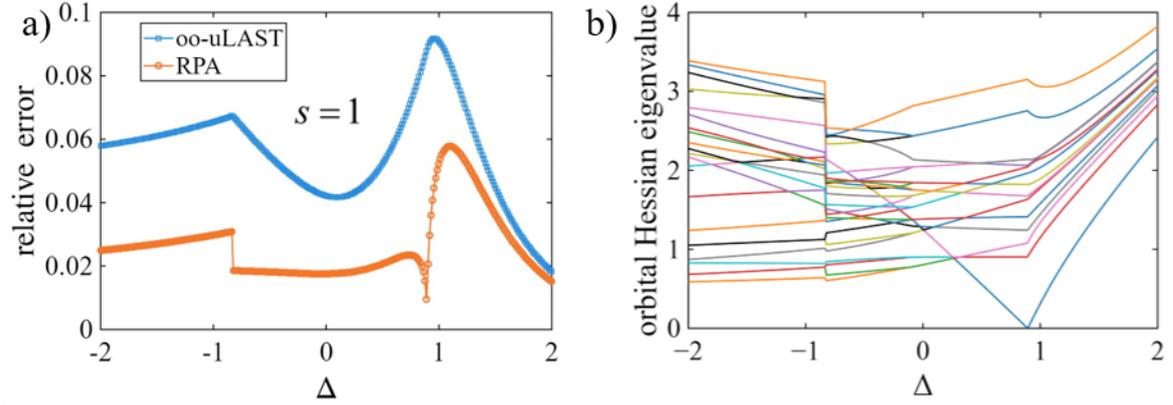

Figure 3: Relation between energetics and mean-field stability diagnostics for an $N=4$ spin-1 ring. a) Relative ground-state energy errors versus $\Delta$ for oo-uLAST and oo-uLAST+RPA. b) Orbital-Hessian eigenvalues of the oo-uLAST references. The softening of a mode near $\Delta \approx 1$ indicates a change in the mean-field reference.

Figure 4 extends the $N=4$ XXZ ring benchmarks to on-site spins $s=\frac{3}{2}, 2, \frac{5}{2}, 3$, where the same qualitative pattern emerges: the RPA step provides a systematic reduction of the mean-field error over a broad parameter range. A common feature is a sharp structure in the vicinity of $\Delta \approx 1$, which coincides with a change in the character of the mean-field solution along the scan (in all cases, a single orbital-Hessian eigenvalue approaches zero, data not shown).

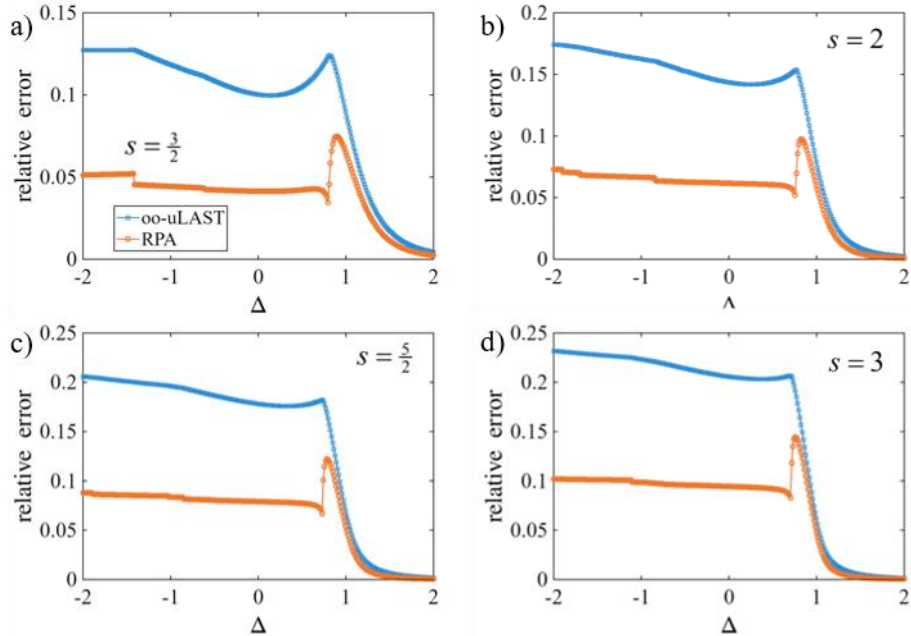

Figure 4: Relative energy errors for an $N=4$ XXZ ring with on-site spins $s=\frac{3}{2}$ (a) through $s=3$ (d).

**$J_1$–$J_2$ model.** Figure 5 benchmarks the isotropic antiferromagnetic $J_1 - J_2$ model at fixed size $N = 24$ for the same geometries as in Figure 2 as a function of the relative magnitude of the frustration-inducing NNN coupling $J_2$. Across all geometries and parameter values, $0 \leq J_2 / J_1 \leq 1$, the oo-uLAST reference (same as in Ref. 14) yields a relative energy error $< 10\%$, with several systems displaying zero error at the Majumdar–Ghosh point $J_2 / J_1 = 0.5$ due to dimer formation.[37] For all systems, the RPA correction yields a systematic error reduction. The gain is especially evident in the $12 \times 2$ PBC lattice, with errors $<1\%$ over a large fraction of the scan region. At the same time, the RPA curves exhibit localized sharp features (spikes, kinks) and occasional sign changes of the error (over-correlation) in narrow windows, most prominently around $J_2 / J_1 \approx 0.6$ for several OBC geometries and around $J_2 / J_1 \approx 0.9$ in some chain/PBC cases, again signaling changes of the optimized mean-field reference, which makes any response-based correction sensitive to small parameter changes. However, the magnitude of the oo-uLAST+RPA error remains substantially below the underlying oo-uLAST reference in most cases. Overall, PBC and more compact clusters generally show larger residual errors, consistent with the increased correlation demands for a larger number of closed loops.

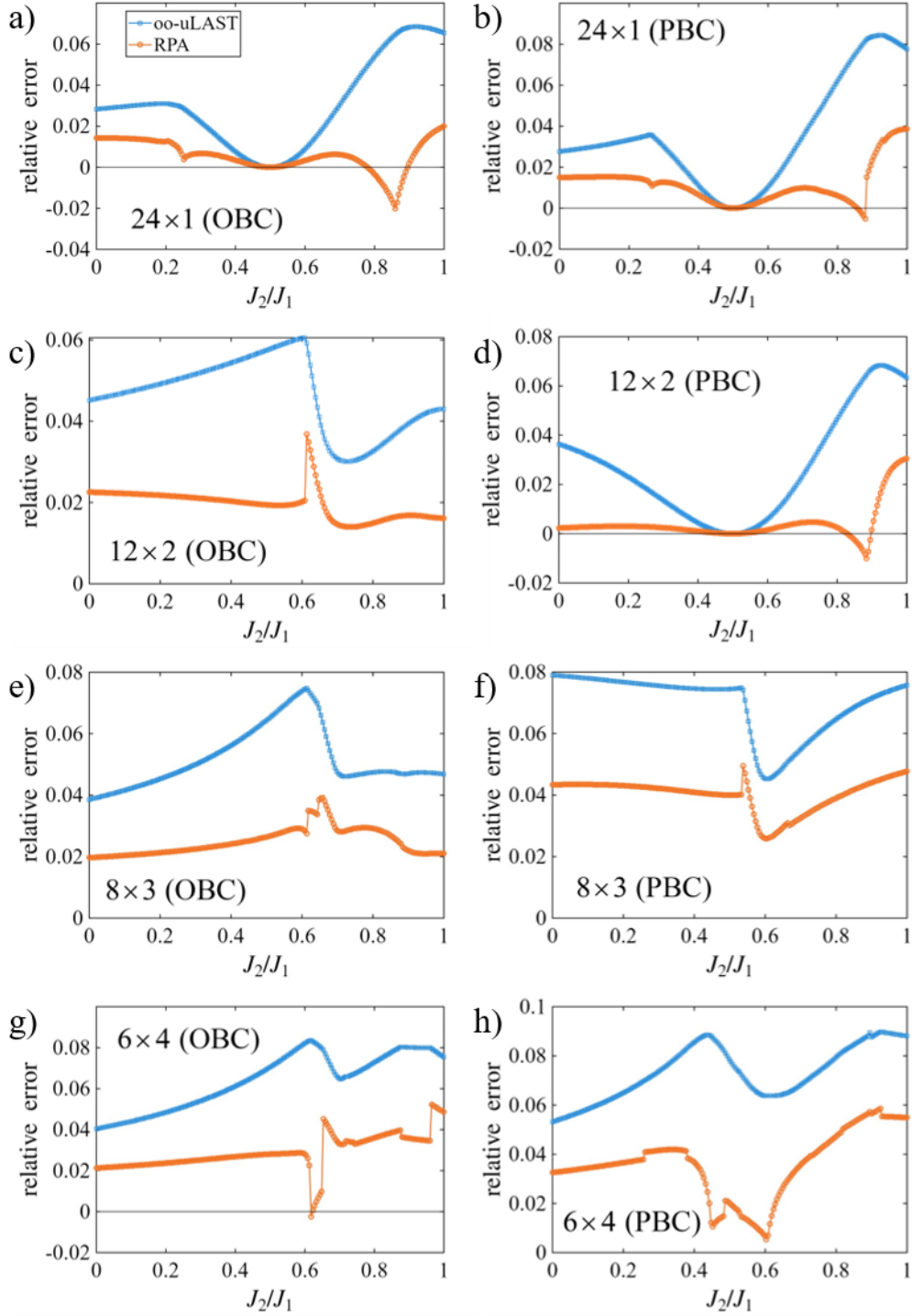

Figure 5: Relative energy errors for the isotropic antiferromagnetic $J_1 - J_2$ Heisenberg model for lattices with 24 spin-1/2 sites (same geometries as in Figure 2).

## 4. Summary and Conclusions

In this work we developed a practical self-consistent field (SCF) and correlation framework for spin Hamiltonians in the fermionic extended Jordan–Wigner (EJW) representation. While mean-field optimization on general coupling topologies can be carried out via Thouless parameters,[14] the complementary SCF formulation, casting the HF energy as a functional of the single-particle density matrix, provides a convenient route to established SCF convergence techniques[16] and a natural interface to response-theory machinery on top of oo-uLAST references, where the free parameters of the EJW transformation and the mean-field state are optimized simultaneously.

We derived the analytic orbital Hessian for EJW-transformed spin Hamiltonians and implemented it as a stability diagnostic for stationary HF solutions. The same building blocks define the corresponding RPA matrix, enabling us to compute ground-state correlation energies (here in the exchange-including ring-CCD convention[21]) at moderate additional cost.

Benchmarks for the *XXZ* model and the antiferromagnetic $J_1 - J_2$ model on one- and two-dimensional geometries show that adding RPA correlation yields a systematic improvement over the underlying mean-field energetics across most parameter regimes. Residual errors tend to increase for more compact clusters and under periodic boundary conditions, consistent with higher correlation demands. Over-correlation by RPA may occur in narrow windows and is typically indicated by a softening of low-lying orbital-Hessian modes due to competing (near-degenerate) mean-field solutions.

Overall, the SCF formulation together with orbital-Hessian diagnostics and an RPA correlation step provides a scalable route to approximate ground-state energies of JW-transformed spin models on arbitrary graphs, and continues a line of work that we consider worthy of further exploration: for spin Hamiltonians, which represent paradigmatic models of strong correlation, fermionic mean-field descriptions based on the JW/EJW-transformation combined with systematically improvable correlation treatments, can offer efficient and accurate approximations.

**Appendix**

**A1. Formulation of the uLAST gradient as a density-matrix functional.** In our earlier work,[14] for the optimization of $\boldsymbol{\theta}$ defining the EJW transformation (see Theory section), we evaluated the gradient based on the Hellmann-Feynman theorem, Eq. (69),

$$g_{kl} \equiv \frac{\partial \langle \Phi | H(\boldsymbol{\theta}) | \Phi \rangle}{\partial \theta_{kl}} = \langle \Phi | \frac{\partial H(\boldsymbol{\theta})}{\partial \theta_{kl}} | \Phi \rangle \;, \tag{69}$$

by computing the expectation value of the derivative of the EJW-transformed Hamiltonian $H(\boldsymbol{\theta})$ with respect to the HF wave function. As an alternative, we here derive a gradient expression that refers only to the density matrix $\boldsymbol{\rho}$ instead of the Slater determinant. This formulation would enable an approach akin to thermal HF, i.e., thermal oo-uLAST. When using the gradient developed here, the oo-uLAST method consists of a nested iterative procedure: a HF routine to obtain a stationary mean-field state followed by gradient-based optimization of the uLAST parameters. Note that the SCF micro-iterations are not required to be solved to a fixed tolerance at all macro-iterations. Instead, one can tighten the SCF stationarity threshold as the optimization proceeds when the uLAST parameters are still far from optimal, while ensuring that the uLAST gradient (derived here for stationary mean-field states) is evaluated in its regime of validity.

The energy gradient, according to Eq. (70), consists of a fixed-density term and a contribution for the coupling with density variations. All partial derivatives are understood to fix the other independent $\theta_{pq}$ angles. We use the notation $\langle \mathbf{A}, \mathbf{B} \rangle \equiv \mathrm{Tr}(\mathbf{A}^\dagger \mathbf{B})$ for the Frobenius product of two matrices in Eq. (70).

$$\frac{\partial E}{\partial \theta_{kl}} = \left( \frac{\partial E}{\partial \theta_{kl}} \right)_{\boldsymbol{\rho}} + \left\langle \frac{\delta E}{\delta \boldsymbol{\rho}}, \frac{\partial \boldsymbol{\rho}}{\partial \theta_{kl}} \right\rangle \tag{70}$$

For a self-consistent solution, with $[\mathbf{F}, \boldsymbol{\rho}] = 0$, $\delta E / \delta \boldsymbol{\rho} = 0$, Eq. (70) simplifies to Eq. (71):

$$\frac{\partial E}{\partial \theta_{kl}} = \left( \frac{\partial E}{\partial \theta_{kl}} \right)_{\boldsymbol{\rho}} . \tag{71}$$

As in the derivation leading to Eq. (28) above, we consider the energy contribution from *xy*-coupling for an ordered pair in Eq. (72):

$$\mathcal{E} = w\mathrm{Tr}(\mathbf{h}\boldsymbol{\rho}_R) = w\mathrm{Tr}(\mathbf{h}\mathbf{R}\mathbf{Z}^{-1}\boldsymbol{\rho}) . \tag{72}$$

$\mathbf{R}$ represents the diagonal matrix of Eq. (73). The relation between $\{\alpha_i\}$ and the parameters $\boldsymbol{\theta}$ will be established in Eq. (81) below.

$$\mathbf{R} = \begin{pmatrix} e^{i\alpha_1} & 0 & 0 \\ 0 & \ddots & 0 \\ 0 & 0 & e^{i\alpha_{N_{\mathrm{orb}}}} \end{pmatrix} \tag{73}$$

Our aim is to compute Eq. (74), with $\boldsymbol{\rho}$ fixed:

$$\frac{\partial \mathcal{E}}{\partial \alpha_q} = \frac{\partial w}{\partial \alpha_q} \text{Tr}(\mathbf{h}\boldsymbol{\rho}_R) + w \text{Tr}\left(\mathbf{h} \frac{\partial \boldsymbol{\rho}_R}{\partial \alpha_q}\right) . \tag{74}$$

In Eq. (75), $\mathbf{E}_{qq} = \mathbf{e}_q \mathbf{e}_q^T$ is a projector along $q$, with $\mathbf{e}_q$ being the respective unit vector.

$$\mathbf{D}_q \equiv \frac{\partial \mathbf{R}}{\partial \alpha_q} = i \mathbf{E}_{qq} \mathbf{R} \tag{75}$$

The consecutive steps in Eq. (76) use: i) the Jacobi formula, ii) Eq. (75), iii) the cyclicity of the trace:

$$\frac{\partial w}{\partial \alpha_q} = w \text{Tr}(\mathbf{Z}^{-1} \boldsymbol{\rho} \mathbf{D}_q) = iw \text{Tr}(\mathbf{Z}^{-1} \boldsymbol{\rho} \mathbf{E}_q \mathbf{R}) = iw \text{Tr}(\mathbf{E}_q \boldsymbol{\rho}_R) = iw (\boldsymbol{\rho}_R)_{qq} . \tag{76}$$

For the derivative of $\boldsymbol{\rho}_R$, we need Eqs. (77) and (78):

$$\frac{\partial \mathbf{Z}}{\partial \alpha_q} = \boldsymbol{\rho} \mathbf{D}_q , \tag{77}$$

$$\frac{\partial \mathbf{Z}^{-1}}{\partial \alpha_q} = -\mathbf{Z}^{-1} \boldsymbol{\rho} \mathbf{D}_q \mathbf{Z}^{-1} . \tag{78}$$

Combining Eqs. (21) and (76) then yields Eq. (79):

$$\frac{\partial \boldsymbol{\rho}_R}{\partial \alpha_q} = \mathbf{D}_q \mathbf{Z}^{-1} \boldsymbol{\rho} - \mathbf{R} \mathbf{Z}^{-1} \boldsymbol{\rho} \mathbf{D}_q \mathbf{Z}^{-1} \boldsymbol{\rho} . \tag{79}$$

The gradient (for fixed $\boldsymbol{\rho}$), Eq. (80),

$$g_q \equiv \frac{\partial E}{\partial \alpha_q} = iw\left[(\boldsymbol{\rho}_R)_{qq} \text{Tr}(\mathbf{h}\boldsymbol{\rho}_R) + (\boldsymbol{\rho}_R \mathbf{h})_{qq} - (\boldsymbol{\rho}_R \mathbf{h} \boldsymbol{\rho}_R)_{qq}\right] , \tag{80}$$

can be formulated as a functional of $\boldsymbol{\rho}$, based on Eqs. (21) und (14).

Reintroducing the *mn* labels, i.e., $\mathcal{E}_{mn} = w_{mn} \text{Tr}(\mathbf{h}_{mn} \boldsymbol{\rho}_R^{mn})$, with $(\mathbf{h}_{mn})_{ij} = \frac{1}{2} J_{mn} \delta_{im} \delta_{jn}$ and $|\Phi_R\rangle = \phi_{m(n)}^\dagger \phi_{n(m)} |\Phi\rangle$, the angles in $\mathbf{R}$ (cf. Eq. (73)) are given by Eq. (81):

$$\alpha_q = (\theta_{mq} - \theta_{nq})(1 - \delta_{mq})(1 - \delta_{nq}) . \tag{81}$$

Employing the chain rule, we get Eq. (82) for the independent parameters $\theta_{kl}$ $(k < l)$.

$$\frac{\partial \mathcal{E}_{mn}}{\partial \theta_{kl}} = \sum_q g_q \frac{\partial \alpha_q}{\partial \theta_{kl}} = \sum_q g_q \frac{\partial(\theta_{mq}-\theta_{nq})}{\partial \theta_{kl}}(1-\delta_{mq})(1-\delta_{nq}) =$$
$$\sum_q g_q [(\delta_{mk}\delta_{ql}+\delta_{ml}\delta_{qk})-(\delta_{nk}\delta_{ql}+\delta_{nl}\delta_{qk})](1-\delta_{mq})(1-\delta_{nq}) = \qquad (82)$$
$$g_l(\delta_{mk}-\delta_{nk})(1-\delta_{ml})(1-\delta_{nl}) + g_k(\delta_{ml}-\delta_{nl})(1-\delta_{mk})(1-\delta_{nk})$$

As $g_m = g_n = 0$, this can be simplified, Eq. (83):

$$\frac{\partial \mathcal{E}_{mn}}{\partial \theta_{kl}} = g_l(\delta_{mk}-\delta_{nk}) + g_k(\delta_{ml}-\delta_{nl}) \ . \qquad (83)$$

The total gradient is a sum over all ordered pairs, Eq. (84):

$$\frac{\partial E}{\partial \theta_{kl}} = \sum_{m \neq n} \frac{\partial \mathcal{E}_{mn}}{\partial \theta_{kl}} \ . \qquad (84)$$

**A2. Fock kernel.** As explained in the main text, to assemble the particle-hole response matrices **A** and **B**, we need the Fock kernel, $K_{pqrs} \equiv \partial F_{pq}/\partial \rho_{rs}$. We first examine the *xy*-contribution, $\mathbf{K}^{xy}$, followed by the simpler *z*-part, $\mathbf{K}^z$, and finally combine both to obtain the total kernel, $\mathbf{K} = \mathbf{K}^{xy} + \mathbf{K}^z$. Starting from the scalar Fock increment **f** for the *xy*-contribution of an ordered pair from Eq. (28), we take the total differential (at fixed uLAST parameters) in Eq. (85).

$$\begin{aligned} d\mathbf{f} &= dw\left[ \mathbf{hRZ}^{-1} - (\mathbf{R}-1)\mathbf{Z}^{-1}\boldsymbol{\rho}\mathbf{hRZ}^{-1} + \mathrm{Tr}(\mathbf{hRZ}^{-1}\boldsymbol{\rho})(\mathbf{R}-1)\mathbf{Z}^{-1} \right] \\ &+ w\left[ \mathbf{hR}d(\mathbf{Z}^{-1}) - (\mathbf{R}-1)d(\mathbf{Z}^{-1})\boldsymbol{\rho}\mathbf{hRZ}^{-1} - (\mathbf{R}-1)\mathbf{Z}^{-1}(d\boldsymbol{\rho})\mathbf{hRZ}^{-1} \right. \\ &\left. -(\mathbf{R}-1)\mathbf{Z}^{-1}\boldsymbol{\rho}\mathbf{hR}d(\mathbf{Z}^{-1}) + \left[ \mathrm{Tr}[\mathbf{hR}d(\mathbf{Z}^{-1})\boldsymbol{\rho}] + \mathrm{Tr}[\mathbf{hRZ}^{-1}(d\boldsymbol{\rho})] \right](\mathbf{R}-1)\mathbf{Z}^{-1} + \right. \\ &\left. \mathrm{Tr}(\mathbf{hRZ}^{-1}\boldsymbol{\rho})(\mathbf{R}-1)d(\mathbf{Z}^{-1}) \right] \end{aligned} \qquad (85)$$

In Eq. (86) we provide the derivative $\kappa_{pqrs} \equiv \partial f_{pq}/\partial \rho_{rs}$, keeping terms in the same order as in Eq. (85).

$$\kappa_{pqrs} = \frac{\partial w}{\partial \rho_{rs}} \left[ \mathbf{hRZ}^{-1} - (\mathbf{R}-1)\mathbf{Z}^{-1}\boldsymbol{\rho}\mathbf{hRZ}^{-1} + \text{Tr}(\mathbf{hRZ}^{-1}\boldsymbol{\rho})(\mathbf{R}-1)\mathbf{Z}^{-1} \right]_{pq}$$

$$+ w \left\{ \left[ \mathbf{hR}\frac{\partial(\mathbf{Z}^{-1})}{\partial \rho_{rs}} \right]_{pq} - \left[ (\mathbf{R}-1)\frac{\partial(\mathbf{Z}^{-1})}{\partial \rho_{rs}}\boldsymbol{\rho}\mathbf{hRZ}^{-1} \right]_{pq} - \left[ (\mathbf{R}-1)\mathbf{Z}^{-1} \right]_{pr} \left[ \mathbf{hRZ}^{-1} \right]_{sq} \right.$$

$$- \left[ (\mathbf{R}-1)\mathbf{Z}^{-1}\boldsymbol{\rho}\mathbf{hR}\frac{\partial(\mathbf{Z}^{-1})}{\partial \rho_{rs}} \right]_{pq} + \left[ \text{Tr}\left( \mathbf{hR}\frac{\partial(\mathbf{Z}^{-1})}{\partial \rho_{rs}}\boldsymbol{\rho} \right) + [\mathbf{hRZ}^{-1}]_{sr} \right] \left[ (\mathbf{R}-1)\mathbf{Z}^{-1} \right]_{pq} + \qquad (86)$$

$$\left. + \text{Tr}(\mathbf{hRZ}^{-1}\boldsymbol{\rho}) \left[ (\mathbf{R}-1)\frac{\partial(\mathbf{Z}^{-1})}{\partial \rho_{rs}} \right]_{pq} \right\}$$

To proceed, we work out the derivatives appearing in Eq. (86). We begin with the overlap derivative in Eq. (87),

$$\frac{\partial w}{\partial \rho_{rs}} = w\text{Tr}\left[ (\mathbf{R}-1)\mathbf{Z}^{-1}\mathbf{E}_{rs} \right] = w\left[ (\mathbf{R}-1)\mathbf{Z}^{-1} \right]_{sr}, \qquad (87)$$

where $(\mathbf{E}_{rs})_{ij} = \delta_{ri}\delta_{sj}$ (in terms of unit vectors along $r$ and $s$, $\mathbf{E}_{rs} = \mathbf{e}_r \mathbf{e}_s^T$); the second equality uses the general relation $\text{Tr}(\mathbf{M}\mathbf{E}_{rs}) = M_{sr}$. For $\mathbf{Z}^{-1}$, we employ Eq. (24) to arrive at Eq. (88).

$$\frac{\partial(\mathbf{Z}^{-1})}{\partial \rho_{rs}} = -\mathbf{Z}^{-1}\mathbf{E}_{rs}(\mathbf{R}-1)\mathbf{Z}^{-1}$$

$$\Leftrightarrow \left( \frac{\partial(\mathbf{Z}^{-1})}{\partial \rho_{rs}} \right)_{pq} = -(\mathbf{Z}^{-1})_{pr}[(\mathbf{R}-1)\mathbf{Z}^{-1}]_{sq} \qquad (88)$$

Substituting Eqs. (87) and (88) into Eq. (86) and using the definitions of Eqs. (89) and (90), yields Eq. (91).

$$\mathbf{Q} = (\mathbf{R}-1)\mathbf{Z}^{-1} \qquad (89)$$

$$\mathbf{V} = \mathbf{hRZ}^{-1} \qquad (90)$$

$$\kappa_{pqrs} = w\left\{ Q_{sr}\left[ \mathbf{V} - \mathbf{Q}\boldsymbol{\rho}\mathbf{V} + \text{Tr}(\mathbf{V}\boldsymbol{\rho})\mathbf{Q} \right]_{pq} \right.$$
$$-V_{pr}Q_{sq} - Q_{pr}V_{sq} + Q_{pr}(\mathbf{Q}\boldsymbol{\rho}\mathbf{V})_{sq} + (\mathbf{Q}\boldsymbol{\rho}\mathbf{V})_{pr}Q_{sq} \qquad (91)$$
$$\left. + \left[ -(\mathbf{Q}\boldsymbol{\rho}\mathbf{V})_{sr} + V_{sr} \right]Q_{pq} - \text{Tr}(\mathbf{V}\boldsymbol{\rho})Q_{pr}Q_{sq} \right\}$$

Eq. (91) represents the kernel increment $\kappa_{pqrs}^{mn}$ for an ordered pair $\langle m,n \rangle$. Summing over all ordered pairs yields the full *xy*-contribution, Eq. (92):

$$\mathbf{K}^{xy} = \sum_{m \neq n} \boldsymbol{\kappa}^{mn} . \qquad (92)$$

The *z*-part, $\mathbf{K}^z$, is straightforward. From Eqs. (8) and (9), we obtain Eq. (93).

$$K^z_{klnm} = \frac{\partial F^z_{kl}}{\partial \rho_{nm}} = [kl \,|\, mn] \qquad (93)$$

The overall kernel, $\mathbf{K}^{AO} = \mathbf{K}^{xy} + \mathbf{K}^z$, is first obtained in the "atomic orbital" (AO) basis (site basis) and then transformed into the MO basis. Let $\mathbf{C} = (\mathbf{v}_1, \mathbf{v}_2, ..., \mathbf{v}_{N_{orb}})$ be the unitary matrix of canonical orbitals (AO-basis eigenvectors of $\mathbf{F}$), with coefficients $(\mathbf{v}_p)_\mu = C_{\mu p}$. Using AO indices, $\mu, \nu, \lambda, \sigma$ and MO indices $p, q, r, s$, the AO → MO transformation of the four-index kernel is given in Eq. (94):

$$K^{MO}_{pqrs} = \sum_{\mu\nu\lambda\sigma} C^*_{\mu p} C_{\nu q} K^{AO}_{\mu\nu\lambda\sigma} C_{\lambda r} C^*_{\sigma s} \,. \qquad (94)$$

In the main text, we drop the MO superscript. The AO → MO transformation preserves the index symmetries, i.e., $K_{pqrs} = K^*_{rspq}$ and $K_{pqrs} = K^*_{qpsr}$.

**Funding.** S.G.T. was supported by the Deutsche Forschungsgemeinschaft (DFG) under Project 535298924. The work at Rice University was supported by the U.S. Department of Energy, Office of Basic Energy Sciences, Computational and Theoretical Chemistry Program under Award DE-SC0001474. G.E.S. acknowledges support as a Welch Foundation Chair (Grant No. C-0036).

**Acknowledgments.** S.G.T. thanks TU Berlin for computational resources.

**Conflicts of Interest.** The authors declare no conflicts of interest.